\documentclass[11pt, dvips]{article}
\def\gsim{\raise0.3ex\hbox{$>$\kern-0.75em\raise-1.1ex\hbox{$\sim$}}}
\def\lsim{\raise0.3ex\hbox{$<$\kern-0.75em\raise-1.1ex\hbox{$\sim$}}}
\begin{document}
\begin{center}
{\LARGE \bf Comments on the recent result of the\\
\vskip 3mm
"Measurement of the neutrino velocity with the \\
\vskip 3mm
OPERA detector in the CNGS beam"}
\end{center}
\vskip 5mm
%  Author List:
\begin{center}
%
%  ***INSTRUCTIONS:***  Replace authors and addresses below with your own:
%
{\bf Luis Gonzalez-Mestres\\}
\vskip 3mm

{\it L.A.P.P., B.P. 110, 74941 Annecy-le-Vieux Cedex, France\\(UMR 5814 CNRS - Universit\'e de Savoie)}
\end{center}
\vskip 5mm
%  Abstract:
\begin{center}
{\large \bf Abstract\\}
\end{center}
\vspace{-0.5ex}
{\bf The recent result by the OPERA experiment, confirming a trend already present in a previous result by MINOS, raises the question of a possible strong violation of standard relativity. In particular, the particles of the standard model would have different critical speeds in vacuum, and such differences would be measurable with nowadays facilities. Although several experimental and phenomenological issues remain open, the situation deserves been studied closely from a theoretical point of view. The data cannot be explained by conventional extrapolations of Planck-scale Lorentz symmetry violation (LSV) patterns. But, as already stressed in our previous papers since 1995, a weak mixing of standard particles with superbradyons (particles with positive mass and energy, and a critical speed in vacuum much larger than the speed of light) can explain such an effect. Superbradyons can be the ultimate constituents of matter (superluminal preons), but they may simultaneously exist in our Universe as free particles just as light can cross a transparent material medium. In this case, a direct mixing between superbradyons and the particles of the standard model would be unavoidable. After briefly examining the experimental situation and the problems it may raise, we comment on the possibility of a superbradyonic mixing, as well as on the implications of a spinorial description of space-time where space translations would form a SU(2) compact group.}

\section{Introduction}

After the MINOS data \cite{MINOS}, OPERA \cite{OPERA} seems to have confirmed that the critical speed of neutrinos in our physical vacuum can be larger than the speed of light. Independently of the controversies \cite{FARGION} that the strong OPERA result may raise, and of what may be the actual value of realistic error bars, there seems to be an experimental trend favouring a critical neutrino speed in vacuum larger that the speed of light. 

We present here a preliminary discussion on the implications of this trend and on the possible dynamical origin of such a phenomenon. Before entering the main subjects covered here (superbradyons and space-time geometry), some preliminary remarks seem necessary.  

Concerning the neutrinos expected from gamma-ray bursts, it must be noticed that in the 100 TeV energy range and even far below this energy, a neutrino critical speed significantly faster than those of light, electrons and other sectors of standard matter would imply spontaneous decays of the neutrino by emitting electron-positron pairs and other particles. Furthermore, as the extra energy required for a pion to decay into a neutrino and a charged lepton can be provided only by the incoming pion mass term, the pion decay may turn out to be impossible. The relevant numerical estimate for such processes is the comparison between the relative difference in critical speed times the energy scale, and the high-energy mass terms of the particles involved. More precisely :

- Taking $c_{\pi}$ to be the pion critical speed in vacuum and $c_{\nu}$ that of the neutrino, a relative difference ($c_{\nu}$ - $c_{\pi}$) $c^{-1}$ $\approx $ 2 x $10^{-5}$ ($c$ = speed of light) would imply the impossibility for on-shell pions to emit neutrinos with energies above $\approx $ 20 GeV, a bound to be compared with the energies considered by OPERA. If, instead, the pion has a critical speed anomaly similar to that of the neutrino (taking the speed of light as the reference critical speed), the effect will naturally propagate to all hadrons and produce observable signatures in cosmic-ray physics and high-energy accelerator experiments. This subject will be dealt with elsewhere but, just to give an example, a 500 GeV proton with ($c_{p}$ - $c$) $c^{-1}$ $\approx ~ 10^{-5}$ ($c_{p}$ = proton critical speed) would spontaneously decay by emitting a photon. 

- Another possibility would be that the sign of the critical speed anomaly varies with the particle considered in a somewhat random way. Even so, phenomenological consistency issues remain. For instance, if ($c_{\pi}$ - $c$) $c^{-1}$ $\approx $ 2 x $10^{-5}$ and baryons have no critical speed anomaly, a 400 GeV charged pion can decay into a baryon - antibaryon pair. 

- Similar considerations apply to neutrino production at astrophysical sources. On-shell pions with ($c_{\nu}$ - $c_{\pi}$) $c^{-1}$ $\approx $ 2 x $10^{-5}$ could not produce the relevant neutrinos above $\approx $ 20 GeV. $\approx $ 100 TeV neutrinos can be produced by on-shell pions only if ($c_{\nu}$ - $c_{\pi}$) $c^{-1}$ is less than $\approx ~ 10^{-12}$. 

A discussion of possible time-of-flight phenomenology for neutrinos from gamma-ray bursts in LSV patterns can be found in \cite{Piran2006}. The existence of spontaneous ("Cherenkow") decays in vacuum for superluminal particles with positive mass and energy was first pointed out in our 1995-96 papers (see, for instance, \cite{Gonzalez-Mestres1995a,Gonzalez-Mestres1996b}). The "Cherenkow" decay of a $\approx ~100 $ TeV neutrino emitting a $e^+ ~e^-$ pair can be avoided only if the difference in critical speed between the electron and the neutrino is less than $\approx ~10^{-16}~c$. This seems to be the strongest tool available concerning the possible effects of LSV for such neutrinos.

\section{Superbradyon mixing}

The possible existence of sectors of matter with a critical speed in vacuum much larger than the speed of light (superbradyonic sectors) was first suggested in our 1995 (January) Moriond article \cite{Gonzalez-Mestres1995a}, and further considered in subsequent papers \cite{Gonzalez-Mestres2010a}. 

Mixing of superbradyons with standard particles \cite{Gonzalez-Mestres1995a,Gonzalez-Mestres1997a}, as well as their detectability at accelerator energies \cite{Gonzalez-Mestres1997b} and their possible presence in our Universe as a component of dark matter \cite{Gonzalez-Mestres1996a,Gonzalez-Mestres2009a} have been addressed, as well as possible cosmological implications \cite{Gonzalez-Mestres2010a,Gonzalez-Mestres2009a,Gonzalez-Mestres1995b}. As just stressed, effects related to "Cherenkow" radiation in vacuum due to the difference in critical speed were already considered in \cite{Gonzalez-Mestres1995a,Gonzalez-Mestres1996b} and in all our papers of the same period.

If the superbradyon kinematics is of the Lorentz type with $c_s$ playing the role of the critical speed, the energy $E_s$ and momentum $p_s$ of a free superbradyon in the vacuum rest frame (the privileged local cosmic inertial frame, VRF) would be given by \cite{Gonzalez-Mestres1995a}: 
\begin{equation}
E_s~=~c_s~(p_s^2~+~m_s^2 ~c_s^2)^{1/2}
\end{equation}
\begin{equation}
p_s~=~m_s~v_s~(1 ~-~v_s^2~c_s^{-2})^{-1/2}
\end{equation}
where $m_s$ is the superbradyon inertial mass and $v_s$ its speed. 

Free superbradyons may actually undergo refraction in the physical vacuum of our Universe (like photons in condensed matter) or exist in it only as quasiparticles and other forms of excitations. They could also be confined, or be able to quit and enter this Universe \cite{Gonzalez-Mestres2010b}. Then, the kinematics and critical speed of superbradyons in our vacuum would not be the same as in an "absolute" vacuum, if such a vacuum exists somewhere.

If massless free superbradyons can exist in our Universe, they are in principle able to mix with the massless particles of the standard model, even if we expect this mixing to be very weak at low energy. As there is no reason for such a mixing to be the same for all particles, this would naturally lead to differences in critical speed between "conventional" particles.

It must be noticed, however, that the mixing with superbradyons does not necessarily produce a higher critical speed. The contrary may also happen if the mixing is "repulsive" in the energy variable. A rough, simplified, exemple would be to consider a conventional massless particle with critical speed $c$ and a massless superbradyon with critical speed $c_s$ and the same momentum, $p$, with a mixing through the energy matrix :
\begin{equation}
H~ =~p~ c_s ~\left( \begin{array}{cc}
\ \epsilon & \rho \\
\rho & 1 \end{array} \right) 
\end{equation}
\noindent
where $\epsilon ~= ~ c~ c_s ^{-1}$ and $\rho $ is a real mixing parameter. Then, requiring for the energy eigenvalues that the determinant of the matrix $~\left( \begin{array}{cc}
\ \lambda - \epsilon & \rho \\
\rho & \lambda - 1 \end{array} \right)$ vanishes, leads to :
\begin{equation}
(\lambda  ~- ~\epsilon) ~(\lambda ~- ~1)~= ~-~\rho ^2
\end{equation}
\noindent
which implies a negative value for $\lambda ~- ~\epsilon $ :
\begin{equation}
(\lambda  ~- ~\epsilon) ~\simeq ~\rho ^2
\end{equation}
with a negative energy shift :
\begin{equation}
\Delta E ~ =~ p ~c ~(\lambda  ~\epsilon ^{-1}~- 1)~\simeq - ~p~ c_s ~\rho ^2
\end{equation}
when $\lambda $ corresponds to the eigenvalue associated to a critical speed close to $c$. Then, the standard particle will in this case emerge with a critical speed below $c$ as a result of its superbradyonic mixing. The values $(\lambda  ~\epsilon ^{-1}~- 1)~\sim ~10^{-5}$ and $c~ c_s ^{-1}~\sim ~10^{-6}$ correspond to $\rho ^2~\sim ~10^{-11}$. 

If a similar mixing is considered with a superbradyonic hole in our vacuum, the sign of $\lambda - ~\epsilon $ will change and the critical speed of the standard particle will become larger than its initial value. More generally, the mixing with superbradyons or superbradyonic objects can lead to positive or negative differences between the critical speed of the original standard particle and that of the corresponding energy eigenstate when the mixing is taken into account.

Thus, a difference in critical speed between neutrinos and light can be due to a superbradyonic mixing for the neutrino increasing its critical speed or to a similar mixing for the photon producing the opposite result, or even to both phenomena simultaneously or to a difference in mixing strength.

Another possibility would be that differences in critical speed between standard particles directly result from their composite structure in terms of superbradyons playing the role of superluminal preons \cite{Gonzalez-Mestres2009a,Gonzalez-Mestres2009b,Gonzalez-Mestres2009c}. If different particles correspond to slightly different structures in the primordial superbradyonic matter, small differences in critical speed can be generated in this way. Such a phenomenon may present the appearance of a symmetry breaking at energies far below Planck scale, and be even comparable to a possible lack of universality of the value of the "elementary" electric charge for standard free particles.

\section{On the possible role of space-time geometry}

In a our 1996 - 1997 papers \cite{Gonzalez-Mestres1996b,Gonzalez-Mestres1997c}, we proposed the use of a spinorial description of space-time where a point of the four-dimensional history our Universe would be described by the two complex components of a SU(2) spinor. The cosmic time $t$ would then be given by the spinor modulus, $t ~= ~\mid \xi \mid $ where $\xi $ is the space-time spinor, and the space at this cosmic time would be described by the corresponding hypersphere in the associated real four-dimensional space. Space translations correspond to SU(2) transformations acting on the spinorial representation. One then has a naturally expanding Universe with positive curvature.

As emphasized in the arXiv.org Post Scriptum to \cite{Gonzalez-Mestres2010a}, such a description of space-time automatically yields the so-called Hubble law (actually first formulated by Georges Lema{\^i}tre in 1927 \cite{Lemaitre1927}), as well as a cosmic redshift. The so-called "Hubble's constant" would then be equal to the inverse of the age of the Universe. Such results are obtained before introducing any reference to matter and its interactions, gravitation or relativity. This suggests that the main problems of standard cosmology can be avoided using this formulation of space-time geometry as the startpoint. See also the Post Scripta to \cite{Gonzalez-Mestres2009a,Gonzalez-Mestres2009c}.

Cosmologically comoving frames correspond then to straight lines crossing the initial point $\xi ~=~0 $, whereas the straight lines crossing a point $\xi ~\ne ~0$ can describe locally all the inertial frames. 

As stressed in the Post Scriptum to \cite{Gonzalez-Mestres2010a}, using a spinorial space-time also corresponds to the experimental fact that spin-1/2 particles exist in Nature, even if the actual generation of such a half-integer angular momentum may correspond to Physics beyond Planck scale and to an associated pre-Big Bang era. A possible hypothesis can be that the situation be similar for differences in critical speed between the particles of our Universe.

Then, the fact that the space translations in our spinorial space-time form a SU(2) compact group can help to unify space-time transformations and internal symmetries in a mathematical structure for standard particles (as seen at energies below Planck scale) where the speed of light would be somehow similar to the electric charge and to other coupling constants, the particle momentum being closer to an equivalent of standard gauge fields. In such a context, critical speed anomalies may be similar to internal symmetry breaking or to a non-universality of the value of {\it e} (the electric charge unit for charged leptons and hadrons).  

\section{Conclusion}

Obviously, further work is needed to check the experimental reliability of the recent OPERA result. Phenomenological issues also require a much closer study. But the possibility that all standard particles do not have the same critical speed in vacuum, and that such differences cannot be related to the usual extrapolations of LSV from the Planck scale, raises several fundamental questions that will be further dealt with in a forthcoming paper.

\end{document}